\documentstyle[amsmath,amssymb,12pt,fleqn,espcrc1,epsf]{article}

\title{QCD in  Nuclear Collisions\thanks{Talk given at ``Quark Matter 2002'', Nantes, July 17-24.} \thanks{This work is supported in part by the 
Department
of Energy, Grant DE-FG02-94ER40819}}%

\author{A.H. Mueller\address{Department of Physics, 
        Columbia University\\
 New York, New York 10027, USA}}

\begin{document}
\maketitle
\begin{abstract}
The relationship between RHIC and HERA data is explored using the idea of saturation (color glass condensate) as a unifying framework for interpretation.  A description of the early stages of a heavy ion collision is given consistent with the RHIC results.  The relevant QCD dynamics at the various stages of a heavy ion collision, from production of gluons to equilibration, is reviewed.  Some comments on current phenomenology are  given.
\end{abstract}

\section{Introduction}

This lecture is devoted to a general semiquantitative description of the early stages of a heavy ion collision.  This is a rapidly evolving subject and it is not yet clear how good a description can be given in terms of QCD.

In the earliest stages of the collision gluon occupation numbers are estimated to be moderately large suggesting a classical gauge theory description.  This fits in nicely with the saturation picture (color glass condensate) of a high-energy heavy ion wavefunction.  The parameters necessary for such a description seem compatible with an analogous description of small-x and moderate $Q^2$ HERA data.

Various discussions of RHIC data in terms of gluon saturation, which dicussions have much in common, are reviewed.  The general picture appears to be coherent although perhaps not yet compelling. The reader should be warned that many of the numbers quoted in the text are theoretical estimates.  There are large uncertainties present, neverthelelss, I believe the estimates to be reasonable.

\section{General features of the early stages of high energy heavy ion collisions}
\subsection{Gluon number densities at the initial time}

Consider a central collision at RHIC, say at ${\sqrt{s}} = 130 GeV.$  The measured transverse energy is ${dE_T\over dy} \approx 500 GeV$\cite{cox} at central rapidities.  Since some of the initial transverse energy goes into longitudinal flow one certainly has

\begin{displaymath}
{dE_T^{in}\over dy} > 500 GeV
\end{displaymath}

\noindent  for the initial transverse energy distribution, which we assume is predominately in gluons.  Thus one can write

\begin{equation}
{dE_T^{in}\over dy} = \left[{dN_g^{in}\over dy d^3b} \cdot \Delta b_z\right] A_g\cdot K \gtrsim 500 GeV
\end{equation}

\noindent with $b_\perp,$ and $b_z$ the coordinates of a point in the overlaping nuclei and where $\Delta b_z$ is the longitudinal width of the volume occupied by gluons at their time of production, $A_g$ is the area, transverse to the colliding beams, occupied by the gluons and $K$ is a typical gluon transverse momentum.  Taking $\Delta b_z = 1/K$\  and\  $A_g \approx 150 fm^2$ gives

\begin{equation}
{dN_g^{in}\over dy d^3b} \simeq 17 /fm^3
\end{equation}

\noindent and with $<K> \approx 2/3$ (see later)

\begin{equation}
{dN_g^{in}\over dy d^2b} \simeq 5/fm^2
\end{equation}

\noindent where (2) and (3), as well as $<K>\approx 2/3 GeV$ represent average values, averaged over all transverse coordinates.  At the center of the nucleus, $b_\perp = 0,$ the more appropriate value of $K$ is about $3/2$ the average value or $ K \simeq Q_s(b_\perp=0) \simeq 1 GeV,$ with $Q_s$ the gluon saturation momentum, so that

\begin{equation}
{dN_g^{in}\over dy d^3b}\vert_{b_\perp=0} \simeq 25/fm^3,\  {dN_g^{in}\over dy d^2b_\perp}\vert_{b_\perp=0} \simeq 7.5/fm^2.
\end{equation} 

\noindent Equations (2-4) give a reasonable guess as to the initial gluon densities averaged over the production region (Equations (2,3)) and at the center of the production region (Equation (4)).

\subsection {Occupation numbers}

The initial gluon number densities discussed just above can be converted to gluon occupation numbers (phase space densities) by

\begin{equation}
f_g^{in} = {(2\pi)^3\over 2\cdot(N_c^2-1)} {dN_g^{in}\over d^3pd^3b} \simeq {(2\pi)^3\over 2(N_c^2-1)} {dN_g^{in}\over dy d^2b_\perp d^2p_\perp}
\end{equation}

\noindent where we use $\Delta y = \Delta p_z/p_z\simeq \Delta b_z\Delta p_z$ and where the 2 in the denominator refers to the number of spin states available to gluons.  We further make the approximation

\begin{equation}
{dN_g^{in}\over dy d^2b_\perp d^2p_\perp} \simeq {1\over \pi Q_s^2} {dN_g^{in}\over dy d^2b_\perp}.
\end{equation}

\noindent Using (4) and (6) in (5) gives

\begin{equation}
f_g^{in}\simeq 1.5
\end{equation}

\noindent at $b_\perp = 0.$  Thus, it would seem that gluon occupation numbers are moderately large at initial times thus fitting well the picture of strong gluon field dominance at early times in heavy ion collisions.

\subsection{The nuclear wavefunction}

It is convenient to view a head-on (central) ion-ion collision in a frame where one of the ions is at rest and the other carries a large longitudinal momentum.  We view the scattering in a light-cone gauge which furnishes the smoothiest connection between gluon quanta present in the wavefunction and the gluons produced at the time of the collision. Roughly, produced gluons at mid-rapidity are already in the wavefunction of the fast ion and are merely ``freed'' by the collision\cite{ler,Bla}.  The picture is illustrated in Fig.1.  The initial wavefunction has a large occupation number for gluons, $f_g\sim 1/\alpha,$ which results in a high density of produced gluons at the time of production.  The fact that $f_g$ is large in the high-energy ion's wavefunction is one of the features that led McLerran and collaborators to call it a color glass condensate.

\begin{figure}[htb]
\epsfbox[0 0 219 145]{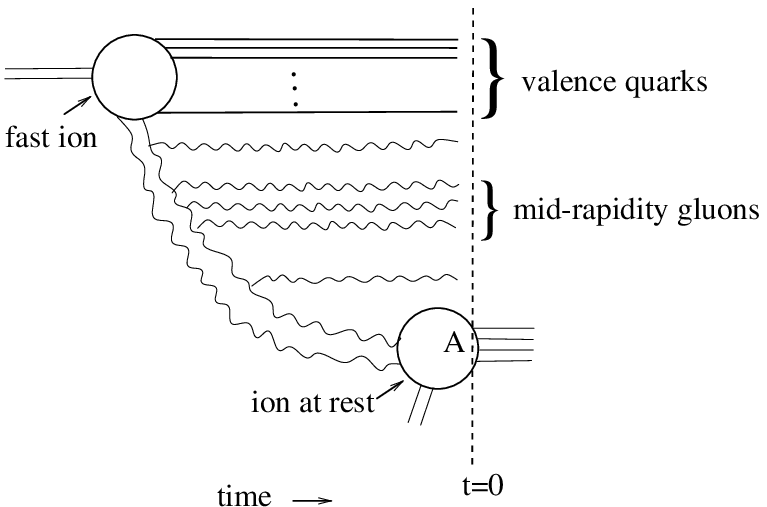}
\caption{}
\end{figure}

\subsection{The McLerran-Venugopalan model}

The McLerran-Venugopalan\cite{Ler} model is a  simple but maybe not unrealistic model of the small-x gluonic part of the wavefunction of a heavy ion at RHIC energies. The Feynman graphs contributing to the wavefunction\cite{Kov} are illustrated in Fig.2. The gluons in the wavefunction correpond to a quasiclassical picture because only tree graphs are involved even when one squares the wavefunction to get the gluon number density.  The gluon number density takes a very simple form\cite{Jal,gov}

\begin{figure}[htb]
\epsfbox[0 0 339 85]{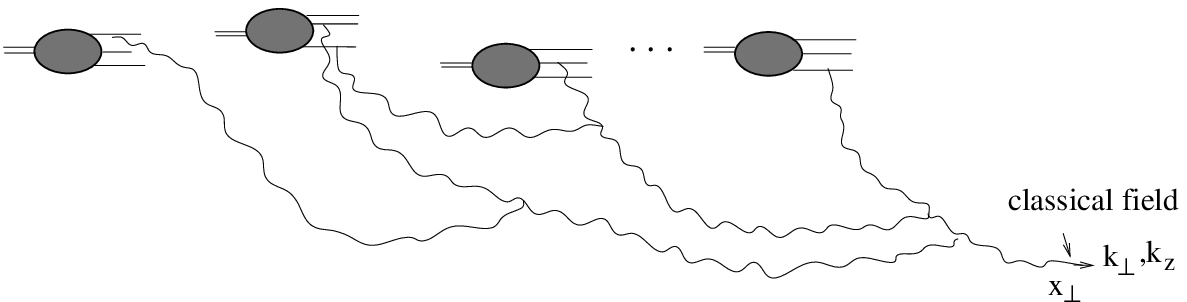}
\caption{}
\end{figure}

\begin{equation}
{dN_g\over dy d^2bd^2k_\perp} \equiv {dxG_A\over d^2bd^2k_\perp} = {N_c^2-1\over 4\pi^4\alpha N_c} \int {d^2x_\perp\over x_\perp^2} e^{-ik_\perp\cdot x_\perp}(1-e^{x_\perp^2Q_s^2/4})
\end{equation}

\noindent where the saturation momentun, $Q_s,$ is given by

\begin{equation}
Q_s^2(b,x) = {4\pi^2\alpha N_c\over N_c^2-1} {2\sqrt{R^2-b^2}}\rho x G_{nucleon}
\end{equation}

\noindent as a function of the impact parameter in the nucleus and of the Bjorken variable, $x.$  Calling

\begin{equation}
{dxG_A\over d^2b} = 2{\sqrt{R^2-b^2}} \rho x G_{nucleon}
\end{equation}

\noindent one can write the saturation momentum as

\begin{equation}
Q_s^2 = {4\pi\alpha N_c\over N_c^2-1} {dxG_A\over d^2b}.
\end{equation}

\noindent It is straightforward to get the gluon occupation number in the wavefunction from (8),  using a formula analogous to (5), and one finds

\begin{equation}
f_g={1\over \alpha N_c} \int {d^2x_\perp\over \pi x_\perp^2} e^{-ik_\perp\cdot x_\perp}(1-e^{-x_\perp^2Q_s^2/4}) \simeq {1\over \alpha N_c} \int_1^\infty {dt\over t} e^{-t k_\perp^2/Q_s^2}
\end{equation}

\noindent where the approximate evaluation, given by the t-integral, is valid when $k_\perp^2/Q_s^2$ is not too large.  This form of the integral neglects a weak $x_\perp^2-$ singularity which is present in $Q_s^2$ via the scale of $x G$ in (9).

From (12) it is apparent that $f_g\sim 1/\alpha$ when $k_\perp^2/Q_s$ is not large.  Occupation numbers of size $1/\alpha$ are characteristic of classical field theory solutions, here using the valence quarks of the nucleons of the nucleus as the sources of color charge. The McLerran-Venugopalan model gives the gluons in the wavefunction of the nucleus as the Weizs\"acker-Williams field generated by the valence quarks of the nucleus.  At RHIC energies this may be a not bad approximation.  At LHC energies quantun evolution in the longitudinal momentum (BFKL evolution) will also be important thus further increasing the gluon numbers by raising the saturation momentum, $Q_s$\cite{Ian,Leo}.  For gluons having $k_\perp/Q_s$ on the order of one, which gluons will dominate the initially produced energy in an ion-ion collision, the occupation number $f_g\sim 1/\alpha.$  The fact that $f_g$ never grows larger than a fixed constant times $[\alpha N_c]^{-1}$ when $k_\perp/Q_s\sim 1$ is the essence of the idea of gluon saturation\cite{Gri,Qiu} and corresponds to the dominance of the initially produced energy density by fields $F_{\mu\nu}\sim 1/g.$

\section{Relationship with HERA phenomenology}
\subsection{The Golec-Biernat W\"usthoff model\cite{nat}}

Many of the properties of the small-x wavefunction of a high-energy heavy ion are also present in high-energy protons and have been extensively studied at HERA.  There is some evidence for saturation at HERA, although most of the evidence is indirect and makes use of models of the Golec-Biernat W\"usthoff type\cite{nat,els,haw,Got,Der}.

In the Golec-Biernat W\"usthoff model one views virtual photon-proton scattering in terms a quark-antiquark dipole, coming from the $\gamma^\ast,$ scattering on the proton.  QCD evolution is put into the proton so that the scattering is that of a bare dipole, having no evolution, colliding with a highly evolved proton wavefunction.  The situation is illustrated in Fig.3 while the formulas are\cite{nat}

\begin{figure}[htb]
\epsfbox[0 0 259 131]{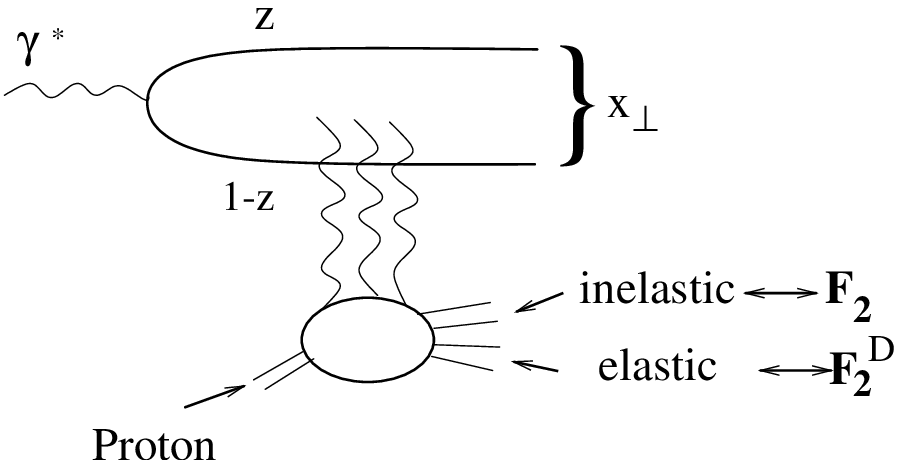}
\caption{}
\end{figure}

\begin{equation}
F_2={Q^2\over 4\pi^2\alpha_{em}}\int d^2x_\perp\int_0^1dz \sum_F e_F^2\vert\psi_T(x_\perp,z,Q)\vert^2\sigma_0(1-e^{x_\perp^2{\bar Q}_s^2/4})
\end{equation} 

\noindent and

\begin{equation}
F_2^D={Q^2\over 4\pi^2\alpha_{em}} \int d^2x_\perp\int_0^1 dz\sum_F e_F^2\vert\psi_T(x_\perp,z,Q)\vert^2{1\over 2} \sigma_0(1-e^{-x_\perp^2{\bar Q}_s^2/4})^2+q{\bar q}g\ {\rm  part}
\end{equation}

\noindent for $F_2$ and the diffractive part of $F_2,$ respectively.

The various factors in (13) have ready interpretations. The ${Q^2\over 4\pi^2\alpha_{em}}$ is the factor connecting $F_2$ to the scattering cross section of an (off-shell) transversely polarized virtual photon on a proton.  $\psi_T$ is the lowest order wavefunction for a photon of virtuality  $Q$  to go into a quark-antiquark pair having transverse coordinate separation $x_\perp$ and with longitudinal momentum fractions  $z$ and $1-z$ for the two members of the dipole.  $e^{-x_\perp^2{\bar Q}_s^2/4}=S$ represents the impact parameter averaged S-matrix for the elastic scattering of a dipole on a proton while $\sigma_0$ represents the size (area) of the proton.

There are three parameters in the model

\begin{equation}
\sigma_0=2.3 mb, x_0=3x10^{-4}\   {\rm and}\ 
   \lambda=0.3
\end{equation}

\noindent where

\begin{equation}
{\bar Q}_s^2=(x_0/x)^\lambda,
\end{equation}

\noindent in units of $GeV^2,$ with ${\bar Q}_s^2$ having the interpretation of an impact parameter averaged quark saturation momentum of the proton where the gluon and quark saturation momenta are related by

\begin{equation}
{\bar Q}_s^2 = {C_F\over N_c} Q_s^2=4/9 Q_s^2.
\end{equation}

The interpretation of (14) goes much the same way.  There is also a term involving the scattering of a quark-antiquark-gluon system on the proton, but this introduces no new parameters.

With the three parameters of the model given as in (15) the G-B W model does a good job in fitting the moderate $Q^2$ and low $x$ HERA data.  Saturation plays an important role because the dipole-proton S-matrix

\begin{equation}
S=e^{x_\perp^2{\bar Q}_s^2/4}
\end{equation}

\noindent becomes zero when $x_\perp\gtrsim 2/Q_s$ thus limiting the contribution of large dipole configurations in (13) and (14) to that allowed by unitarity. This limitation is especially important in diffractive scattering which in perturbation theory has a strong divergence coming from large dipole sizes.

In the HERA range the value of ${\bar Q_s^2}$ from (16) is between 1 and 2 $GeV^2$  giving $Q_s^2\sim (2-4) GeV^2.$  As we shall see later this value may be a little large and a more realistic model can be expected to give somewhat lower values of $Q_s^2.$

\subsection{Going to impact parameter space}

One can get information on the S-matrix for dipole-proton scattering as a function of impact parameter by using $\gamma\ast$   proton $\rightarrow J/\psi$  proton and $\gamma\ast$  proton $\rightarrow \rho$ proton quasielastic scattering data\cite{ier}.  The process is pictured in Fig.4 for the production of longitudinally polarized $\rho$-mesons at momentum transfer $\Delta.$  To analyze the process one also needs a model of the $\rho$-meson wavefunction which is taken from earlier phenomenological analyses\cite{set,aev,res}.  In the approximation that the  dipole-proton scattering amplitude is purely imaginary one can write the amplitude in terms of the square root of the differential cross section.  It is straightforward to arrive at

\begin{figure}
\epsfbox[0 0 208 116]{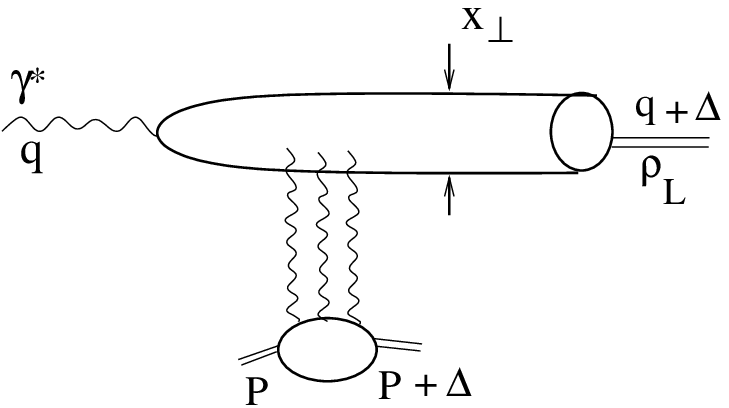}
\caption{}
\end{figure}

\begin{equation}
S(b, x_\perp, x) = 1 - {1\over 2\pi^{3/2}N} \int d^2\Delta e^{-i\Delta\cdot b}{\sqrt{{d\sigma\over dt}}}
\end{equation}

\noindent where

\begin{equation}
N=(\psi_{\rho_L},\psi_{\gamma^\ast})
\end{equation}

\noindent is the overlap of the wavefunctions for a longitudinal virtual photon with a longitudinal $\rho-$meson.  $S$ depends on the impact parameter, $b,$ on the Bjorken $x-$variable, and on the transverse size of the dipole pair, $x_\perp,$ which is fixed, approximately, in terms of $Q.$

In order to interpret (19) in terms quantities we have used in the Golec-Biernat W\"usthoff model  write  $S$  as in (18)\cite{ier}.  (This form for  $S$  is not generally true, but it is not unreasonable when gluon number densities are large.)  From fitting the data at $x = 5x10^{-4}$ and $Q^2=3.5 GeV^2 (x_\perp \simeq 0.4 fm)$ one gets

\begin{equation}
{\bar Q}_s^2(b=0)\simeq 0.5 GeV^2\ \ {\rm  or}\ \  Q_s^2(b=0) \simeq 1.1 GeV^2.
\end{equation}

\noindent Although the values given in (21) have considerable uncertainity, mainly from deciding what value of $x_\perp$ to assoicate with a given $Q^2,$ I think they are reasonable.  They are somewhat smaller than those that come from the Golec-Biernat W\"usthoff model.  Using (11) to write

\begin{equation}
{\bar Q_s^2} (b,x) = {2\pi^2\alpha\over N_c}\  {dxG\over d^2b}
\end{equation}

\noindent we are led to

\begin{equation}
{dxG\over d^2b}\vert_{b=0}\simeq 5/fm^2\ \ {\rm and}\ \ xG\simeq 6
\end{equation}

\noindent for the values of $x$ and $Q^2$ quoted above.

\subsection{From HERA to RHIC}

Now let's be really adventurous and scale (23) back to RHIC energies, $x \simeq 0.01,$ by using the scaling law

\begin{equation}
\bar{Q}_s^2(x) = ({x_0\over x})^{0.3}{\bar Q}_s^2(x_0)
\end{equation}

\noindent found in the Golec-Biernat W\"usthoff analysis.  This gives

\begin{equation}
xG\vert_{RHIC}\simeq 2.5
\end{equation}

\noindent a little larger than most perturbative QCD fits, but not a crazy number at all.

Now that we have $xG$ for the proton at RHIC energies we can use (9) and (10) to get the corresponding gluon density and saturation momentum.  For Au at RHIC energies one finds from (9), (10) and (25)

\begin{equation}
Q_s^2\vert_{b=0}\simeq 1.25 GeV^2 \ \ {\rm and}\ \ {dxG_A\over d^2b}\vert_{b=0}\simeq 5/fm^2
\end{equation}

\noindent not too different from our estimate given in Sec.2.1 for the values of the typical momentum and number density of gluons produced in a heavy ion collision. We shall come back a little later to the question of the relationship between the number density of gluons in the wavefunction of an ion just before collision and the number density of initially produced gluons.

\section{What dynamics can (should) one use?}

For the wavefunction of a heavy ion, and at RHIC energies, the McLerran-Venugopalan model should be adequate.  However, by LHC energies, gluon evolution will certainly become important\cite{Ian,Leo} and may even be significant at RHIC energies.  The density of gluons in the wavefunction (26) and initially produced (2-4) is  so large at  RHIC that the idea of viewing the wavefunction and the production process in terms of hadrons and in terms of nucleon-nucleon collisions, respectively, does not make sense.  The collision must be viewed in terms of quarks and gluons.

The McLerran-Venugopalan model uses classical gluon dynamics with random color sources given by the valence quarks.  Because gluon occupation numbers are reasonably large the freeing (production) of gluons in an ion-ion collision should also be amenable to a description in terms of classical gluon dynamics.  So far there have been two important approaches to this problem.

Krasnitz and Venugopalan\cite{itz,ara} have developed a numerical procedure for calculating the early stages of a heavy ion collision in terms of discretized classical Yang-Mills equations.  The results of these calculations will be discussed shortly. Kovchegov\cite{che} has tried to calculate the gluon production cross section by looking at Feynman graphs. His procedure for gluon production in proton-nucleus and nucleus-proton collisions gives reliable results.  (Nucleus-proton scattering is not {\underline {manifestly}} the same as proton-nucleus scattering in Kovchegov's calculation because of the use of a particular light-cone gauge.)  It is still an open question whether his procedure is valid for nucleus-nucleus collisions, although the results which come out are quite reasonable.

\subsection{Krasnitz, Nara, Venugopalan calculation\cite{ara}}

Starting with a McLerran-Venugopalan wavefunction as an initial condition Krasnitz, Nara and Venugopalan numerically solve discretize Yang-Mills equations as the nuclei in a heavy ion collision overlap.  Gluons appear to be freed on a time scale $\tau \simeq 3/Q_s.$  Just before the collision the distribution of gluons in one of the nuclear wavefunctions is given by (8) and with gluon occupation numbers given by (12).  Just after the gluons are freed the distribution of liberated gluons fits well to the empirical formula

\begin{equation}
{dN_g^{in}\over dy d^2b_\perp d^2p_\perp} = {N_c^2-1\over 4\pi^3} {1\over \alpha N_c}  {0.11\over e^ {{\sqrt{p_\perp^2+m^2}}/T_{eff}}-1}
\end{equation}

\noindent with $T_{eff} \simeq 1.1 Q_s$ and $ m\simeq 0.08 Q_s$ giving

\begin{equation}
f_g^{in} \simeq {1\over \alpha N_c}\ {0.11\over e^{p_\perp/T_{eff}}-1}
\end{equation}

\noindent so long as $p_\perp/T_{eff}$ is not too small.

Comparing (8) to (27) one sees that  $c$, the fraction of gluons freed in the collision, obeys $c\simeq 1/2.$  In the wavefunction $<p_\perp /Q_s> \simeq 0.6$ while in the initial distribution (28), $<p_\perp/Q_s> \simeq 1.5$ so that the freeing process seems to create a lot of additional transverse energy which is not present in the ion wavefunction. One worrisome result of the calculation is that the occupation number, given by (28), is quite small so that classical evolution cannot be expected  to be reliable after the gluons have been freed and one might even worry about the validity of the freeing calculation being done at a classical level.

\subsection{Classical field theory versus kinetic theory}

When $f_g \sim 1/\alpha$ kinetic theory does not apply and classical field theory is the right approach.  Thus in the earliest stages of a heavy ion collision, during the time of the freeing of the gluons and immediately after, classical field theory is the natural dynamics to use.  There may be a practical problem, however, because the Krasnitz-Nara-Venugopalan calculation\cite{ara} suggests that there is a rapid lowering of the occupation number as the gluons are being liberated.  Thus, although there is a $1/\alpha$ factor in (28) the actual value of $f_g$ is not large in any practical circumstance.  If it should turn out that the classical calculation is not valid numerically during the freeing of gluons the correct dynamics would then be a full quantum field theoretic calculation starting from just before the collision up to the time gluons have been freed, a very difficult calculation.

After gluons have been freed and in the regime (perhaps purely a formal regime) where $1<f_g<1/\alpha$ classical field theory and the Boltzmann equation are equivalent procedures for calculating the time evolution of the system\cite{Son}. 

When $f \simeq 1$ classical field theory, which only keeps the $f_g$ in statistical factors $(1+f_g)$ is an underestimate of the interation strength of gluons while the Boltzmann equation is systematic.

The ``bottom-up'' picture of evolution\cite{iff} uses the Boltzmann equation, including elastic and inelastic collision terms, and follows the gluon system from shortly after gluons have been produced up to the time when equilibration occurs.

\subsection{``Bottom-up'' picture\cite{iff};Kharzeev, Levin,Nardi picture\cite{rdi,eev,vin}}

The ``bottom-up'' picture of thermalization applies to a collision of heavy ions where the freed transverse energy is dominated by gluons having transverse momenta well into the perturbative scale. At RHIC energies the transverse momenta which dominate are in the region of one GeV and so one might expect the picture to be qualitatively and even semiquantitatively correct but precise calculations are always going to be hard to come by.  The picture does not require that the initial state of the system have large occupation numbers but we shall describe it in the context of the initial conditions being given by gluon saturated heavy ion-wavefunctions.

Consider a head-on high energy heavy ion collision. Our description will refer to gluons in the central unit of rapidity in the center of mass system.  The time over which the gluons are freed is $\tau \simeq 1/Q_s.$  This is also the time scale on which the higher rapidity gluons separate, in the longitudinal spatial direction, from the gluons in the central unit of rapidity with which we are concerned. We presume that a finite fraction, c, of the gluons in the initial wavefunction have been freed in the collision as discussed in sec.2.3 giving initial gluon occupation numbers on the order of $1/\alpha.$

As $Q_s\tau$ grows the occupation number decreases due to elastic scattering during the longitudinal expansion  of the system. When $f_g$ is significantly less than $1/\alpha$ the Boltzmann equation, the basic dynamics in the bottom-up picture, begins to be applicable.  When $Q_s\tau$ gets as large as $\alpha^{-3/2}$ the occupation number for gluons becomes less  than one.

When $\alpha^{-3/2} < Q_s\tau < \alpha^{-5/2}$ the dominant process is soft gluon emission from the hard gluons.  During this period the density of soft gluons (having $p_\perp \sim {\sqrt{\alpha}}\ Q_s)$ is catching up with the density of hard gluons (having $p_\perp \sim Q_s$), although both densities are decreasing in time due to longitudinal expansion.

When $Q_s\tau \sim \alpha^{-5/2}$ the density of soft gluons reaches that of hard gluons.  At later times two important changes happen.  (i)  The soft gluons stimulate further emission of soft gluons so that now the number density of soft gluons grow with time in spite of the longitudinal expansion. (ii)The soft gluons equilibrate amongst themselves with the hard gluons serving as a source of energy causing the temperature of the soft gluons to grow with  time as $T\sim \alpha^3Q_s\tau.$  The rate of flow of energy from hard gluons continues to grow until, at $Q_s\tau \sim \alpha^{-13/5},$  the hard gluons disappear, having lost all their energy, and one is left with a completely equilibrated system.  At later times the temperature decreases $\tau^{-1/3}$ as the equilibrated system continues to expand. The various stages in the equilibration process are shwon in Fig.5.  Finally, it is important to note that it is probably overly optimistic to believe that these separate regimes are sharply distinct at RHIC energies.

\begin{figure}
\epsfbox[0 0 380 191]{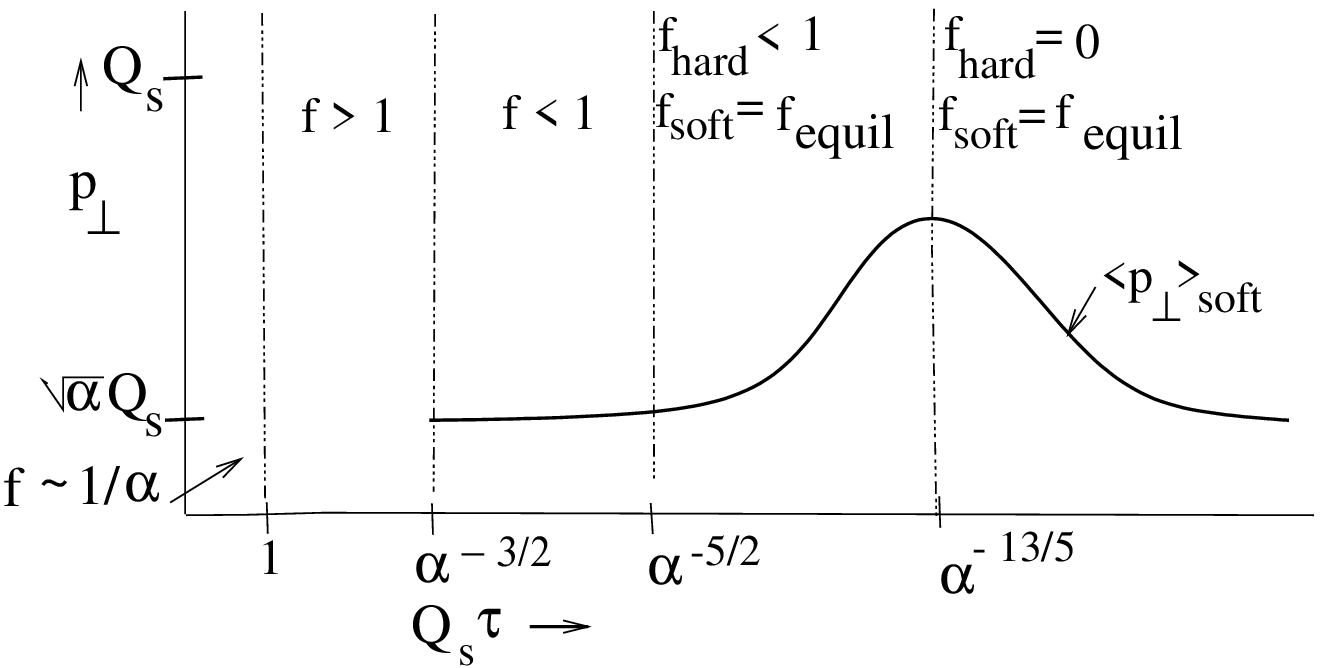}
\caption{}
\end{figure}

Now let us try to simply characterize charged hadron production at RHIC.  Take a central collision.  Then at mid-rapidity the initial gluon density in the McLerran-Venugopalan model is\cite{rdi}

\begin{equation}
{dN_g^{in}\over dyd^2b} = c 2{\sqrt{R^2-b^2}}\ \rho x G
\end{equation}

\noindent where  $c$  is the fraction of gluons liberated. Eq.(29) is easily obtained by integrating (8) over $k_\perp,$ using (9), and inserting the gluon freeing factor $ c.$  Now at late times the experiments measure the rapidity distribution of charged hadrons.  From (29) we get

\begin{equation}
<{2\over N_{part}}\ {dN_{ch}\over dy}> = {2\over 3} Rc  xG (x,<Q_s^2>) \simeq 3.8
\end{equation}

\noindent where $N_{part}$ is the number of participants, the factor ${2/3}$ accounts for the fraction of charged particles and  R  is an inelasticity factor as the partonic  system evolves to equilibrium.  (R also accounts for any, presumably small,  change in particle numbers as quarks and gluons go to hadrons.)  The experimental value\cite{Bac} of 3.8 corresponds to ${\sqrt{s}} = 200GeV.$

Kharzeev, Levin and Nardi take a very simple picture where $R=1$ so that ${2\over 3} c x G\simeq 3.8.$  The KLN analysis was the first detailed analysis of particle production at RHIC using the saturation (color glass condensate) picture and so it has great historical importance.  I believe it is probably difficult to get $c x G$ as large as 5.7.

The Krasnitz, Nara, Venugopalan picture\cite{ara} is also a saturation picture where, however, one of the primary achievements was to calculate $c$ and find that $c\simeq 1/2$ at RHIC energies. Moderately large values of  $R$ and $xG$ are necessary to agree with the data.

The ``bottom-up'' picture does not give a precise value for  $R,$  and of course there is no calculation of $c$ or of $xG.$  However, given values of  $R, c$ and $xG$ one does get values (estimates) for the time at which equilibration occurs and the temperature at that time.  For $R\simeq 3, x \simeq 1$ and $xG\simeq 2$ one gets $T_{equil}\simeq 230 MeV$ and $\tau_{equil}\simeq 3.6 fm,$ values which seem reasonable.

The KLN, KNV and ``bottom-up'' pictures are all based on the same ideas, and differ only in the focus of the calculation and in the various details.

\section{Surprises, puzzles, connections}
\subsection{Gluon liberation, proton-nucleus vs nucleus-nucleus}

Consider gluon emission in quark-nucleus scattering where the coherence time of the gluon is larger than the nuclear radius. The process  is best viewed in the rest system of the nucleus and is shown in Fig.6 in lowest order.  The graph in part (a) of the figure corresponds to a gluon in the wavefunction of the quark as the quark reaches the nucleus.  Graph (b) corresponds to gluon emission after the quark passes the nucleus.  The calculation is straightforward to perform\cite{gov}.  The square of graph (a) liberates all gluons having $k_\perp^2/Q_s^2 \lesssim 1$ while the square of graph (b) gives the ``shadow'' of graph (a) producing gluons with a $dk_\perp^2/k_\perp^2$ spectrum for $k_\perp^2/Q_s^2 \lesssim 1.$  The interference terms do not contribute when $k_\perp^2/Q_s^2\lesssim 1$ and they cancel the direct terms for $k_\perp^2/Q_s^2 \lesssim 1.$  The calculation can be generalize to an incident proton on a nucleus with a similar result.  Thus one gets

\begin{figure}
\epsfbox[0 0 405 105]{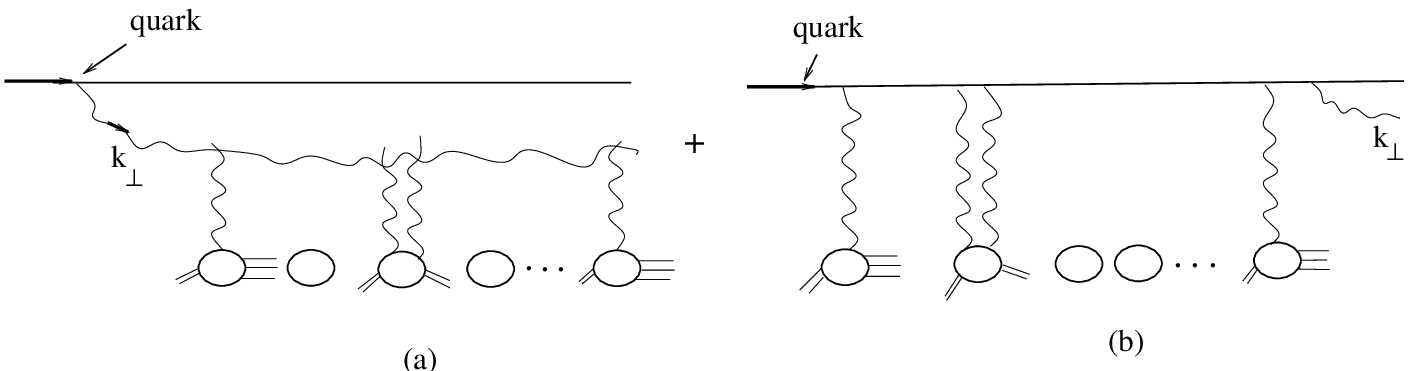}
\caption{}
\end{figure}

\begin{equation}
{dN_g\over dy} = 2 x G_{proton}(x, Q_s^2)
\end{equation}

\noindent corresponding to a gluon liberation factor of 2. The fact that the gluon liberation factor is 2 rather than 1 corresponds to the ``shadow'' term and has a quantum mechanical interpretation analogous to the result of $\sigma = 2\cdot \pi R^2$ for  the scattering of a particle off a hard sphere in non-relativisstic quantum mechanics.

Now, what is surprising is that the Krasnitz, Nara, Venugopalan calculation gives a much smaller number, $c \approx 1/2.$  The lack of a ``shadow'' term in heavy ion collisions is perhaps natural.  Then we are left with trying to understand why  $c$ is near to $1/2$ rather than 1.  In the case of proton-A scattering most of the freed gluons coming from the (a) term in Fig.6 have a momentum, $k_\perp,$ in the proton's wavefunction obeying $k_\perp/Q_s \ll 1,$ parametrically for large $Q_s.$  For A-A collisions the freed gluons in the wavefunction have $k_\perp/Q_s$ on the order of one so perhaps it is not {\underline{so}}\ surprising that some of the gluons having $k_\perp$ on the order of $Q_s$ are not freed.  This indeed, might be expected if it were not for the fact that the freeing process, as described in Sec.4.1 creates so much transverse momentum.
 
\subsection{$v_2(p_\perp)$}

One of the strongest indications of an early equilibration of the quark gluon plasma in ion-ion collisions at RHIC is elliiptic flow\cite{ult}.  Detailed calculations using the equations of ideal hydrodynamics suggest that thermalization must occur by $\tau \simeq 0.6 fm$ in order to get a sufficiently large elliptic flow\cite{nen,yak}.  On the other hand it appears extremely difficult to get such early equilibration from perturbative QCD; the interaction strength does not seem strong enough to cause equibibration at such early times\cite{eau,lan,ssy,nar,ass}.  One can see the difficulty very simply. To match RHIC data the thermalized quark-gluon plasma at $\tau \approx 0.6$ has a temperature about $330 MeV$ corresponding to a typical quark and gluon momentum of about $1 GeV.$  Perturbative calculations should be reliable to test whether gluons having $2GeV$ can be in equilibrium.  The mean free path for a $2GeV$ gluon to have a collision with momentum transfer $2GeV$ is on the order of 10fm at $T=330 MeV.$  Thus it is extremely difficult to see how 2 GeV gluons can be thermalized at $\tau \simeq 0.6 fm.$  On the other hand gluons of $500 MeV$ are still in the process of being produced at $\tau\simeq 0.6 fm.$  Indeed, Krasnitz and Venugopalan\cite{itz} find that the dominant energy density only becomes produced when $\tau \simeq 3/Q_s.$  Thus, while it is not impossible that an effective thermalization of gluons in the regime $0.5-1 GeV$ could occur at $\tau \simeq 0.6 fm$ it would be puzzling from the QCD dynamics point of view.

\subsection{Saturation and $p_\perp-$broadening}

Experiments on $\mu-$pair production in proton-A reactions observed a broadening in the transverse momentum, $p_\perp,$ spectrum of the $\mu-$pair\cite{alde}.  On the theoretical side one expects the change in the width of the $p_\perp^2-$spectrum to be\cite{Pei}

\begin{equation}
\Delta p_\perp^2 = {2\pi^2\alpha\over N_c} L \rho x G
\end{equation}

\noindent where L  is the length of nuclear material the annihilating antiquark passes through, $\rho$ is the nuclear density, and $x G$ is the nucleon gluon distribution. With $xG\simeq 1$ (32) is in reasonable agreement with data.  Now recalling (9) and (17) we see that (9) and (32) are really the same formula\cite{aie}.  Thus we can get another formula for $Q_s^2$ as

\begin{equation}
Q_s^2(b) = {N_c\over C_F} \cdot 2{\sqrt{R^2-b^2}}\ \cdot {d\Delta p_\perp^2\over dz}.
\end{equation}

\noindent Now take the experimental number

\begin{equation}
{d\Delta p_\perp^2\over dz} \simeq 0.02 GeV^2/fm
\end{equation} 

\noindent for the broadening of the $\mu-$pair spectrum at fixed target energies.  This gives

\begin{displaymath}
Q_s^2\vert_{b=0} \simeq 0.6 GeV^2.
\end{displaymath}

\noindent This number is about a factor of 2 smaller than what one might expect at RHIC, but at RHIC energies ${d\Delta p_\perp^2\over dz}$  may well be significantly larger than at fixed target energies.  A comparison of the $\mu-$pair spectrum of proton-proton and proton-nucleus reactions could give a good direct measurement of ${\bar Q}_s^2$ while a comparison of the $J/\psi$ production transverse momentum spectrum could give a good direct measurement of $Q_s^2.$


\begin{thebibliography}{9}
\bibitem{cox} PHENIX collaboration, K. Adcox et al., Phys. Rev. Lett. {\bf 87} (2001) 052301.
\bibitem{ler}	A.H. Mueller, Nucl.Phys. {\bf B572} (2000) 227.
\bibitem{Bla} J.-P. Blaizot and A.H. Mueller, Nucl. Phys. {\bf B289} (1987) 747.
\bibitem{Ler} L. McLerran and R. Venugopalan, Phys. Rev. {\bf D49} (1994) 2233; {\bf D49} (1994) 3352; {\bf D50} (1994) 2225.
\bibitem{Kov}	Yu. V. Kovchegov, Phys. Rev.{\bf D55}(1997) 5445. 
\bibitem{Jal} J. Jalilian-Marian, A. Kovner, L. McLerran and H. Weigert, Phys. Rev.{\bf D55} (1997) 5414.
\bibitem{gov} Yu.V. Kovchegov and A.H. Mueller, Nucl. Phys. {\bf B529} (1998) 451.
\bibitem{Ian}	E. Iancu, A. Leonidov and L. McLerran, hep-ph/0202270.
\bibitem{Leo}	E. Iancu, A. Leonidov and L. McLerran, Nucl. Phys. {\bf A692} (2001) 583; E. Ferreiro, E. Iancu, A. Leonidov and L. McLerran, hep-ph/0109115. 
\bibitem{Gri}	L.V. Gribov, E.M. Levin and M.G. Ryskin, Phys. Rep.{\bf 100} (1983)1.
\bibitem{Qiu}	A.H. Mueller and J. Qiu, Nucl.Phys.{\bf B268} (1986) 427.
\bibitem{nat}	K. Golec-Biernat and M. W\"usthoff, Phys. Rev.{\bf D59} (1999) 014017; Phys. Rev. {\bf D60} (1999) 114023.
\bibitem{els}J. Bartels, K. Golec-Biernat and H. Kowalski, hep-ph/0203258.
\bibitem{haw}	J.R. Forshaw, G. Kerley and G. Shaw, Phys. Rev.{\bf D60} (1999) 074012; Nucl. Phys. {\bf A675} (2000) 80. 
\bibitem{Got}	E. Gotsman, E.M. Levin, U. Maor and E. Naftali, Eur.Phys.J. {\bf C10} (1999) 689.
\bibitem{Der} M.Mc Dermott, L. Frankfurt, V. Guzey and M. Strikman, Eur.Phys. J. {\bf C16} (2000) 641.
\bibitem{ier}	S. Munier, A.M. Sta\'sto and A.H. Mueller, Nucl. Phys. {\bf B603} (2001) 427. 
\bibitem{set}	H.G. Dosch, T. Gousset, G. Kulzinger and H. J. Pirner, Phys. Rev.{\bf D55} (1997) 2602; H.G. Dosch, G. Kulzinger and H. J. Pirner, Eur.Phys.J. {\bf C7} (1991) 73.
\bibitem{aev}	J. Nemchik, N.N. Nikolaev and  B.G. Zakharov, Phys. Lett.{\bf B341} (1994) 228,  J. Nemchik, N.N. Nikolaev, E. Predazzi, and B.G. Zakharaov, Z. Phys. {\bf C75} (1997) 71.
\bibitem{res} A. Caldwell and M.S. Soares, Nucl.Phys. {\bf A696} (2001) 125.
\bibitem{itz}	A. Krasnitz and R. Venugopalan, Nucl. Phys. {\bf B557} (1999) 537; Phys. Rev.Lett.{\bf 86} (2001) 1717.
\bibitem{ara}	A. Krasnitz, Y. Nara and R. Venugopalan, Phys. Rev. Lett.87 (2001) 192302-1.
\bibitem{che}	Yu. V. Kovchegov, Nucl.Phys. {\bf A692}(2001) 557.
\bibitem{Son}	A.H. Mueller and D. Son (unpublished).
\bibitem{iff}R. Baier, A.H. Mueller, D. Schiff and D.T. Son, Phys. Lett. {\bf B502} (2001) 51; hep-ph/0204211.	
\bibitem{rdi}D. Kharzeev and M. Nardi,Phys. Lett. {\bf B507} (2001) 121.
\bibitem{eev} D. Kharzeev and E. Levin, Phys. Lett. {\bf B523} (2001) 79.
\bibitem{vin}	D. Kharzeev, E. Levin and M. Nardi, hep-ph/0111315.
\bibitem{Bac}	PHOBOS collaboration, B.B. Back et al.Phys.Rev.Lett. {\bf 88}(2002) 022302.  
\bibitem{ult}J.-Y. Ollitrault, Phys. Rev. {\bf D46} (1992) 229.	
\bibitem{nen}P.F. Kolb, U.W. Heinz, P. Huovinen, K.J. Eskola and K. Tuominen, Nucl. Phys. {\bf A696} (2001) 197; P. Huovinen, P.F. Kolb, U.W. Heinz, P.V. Ruuskanen and S.A. Voloshin, Phys. Lett. {\bf B503} (2001) 58, P.F. Kolb, P. Huovinen, U.W. Heinz and H. Heiselberg, Phys. Lett. {\bf 500} 232.
\bibitem{yak}D. Teaney, J. Lauret and E.V. Shuryak, Phys. Lett. {\bf 86} (2001) 4783.	
\bibitem{eau}J. Serreau and  D. Schiff, JHEP {\bf 0111} (2001) 039.	
\bibitem{lan}J. Bjoraker and R. Venugopalan, Phys. Rev.{\bf C63} (2001) 024609.	
\bibitem{ssy}A. Dumitru and M. Gyulassy, Phys. Lett. {\bf B494} (2000) 215.
\bibitem{nar}D. Molnar and M. Gyulassy, Nucl. Phys. {\bf A697} (2002) 495.
\bibitem{ass}S. Bass (talk at this conference).
\bibitem{alde}D.M. Alde, et al., Phys. Rev. Lett. {\bf 66} (1991) 2285, and references therein.
\bibitem{Pei} R. Baier, Yu. L. Dokshitzer, A.H. Mueller, S. Peign\'e  and D. Schiff, Nucl. Phys. {\bf B484} (1997) 265.	
\bibitem{aie}R. Baier (Plenary talk at a this conference).
\end{thebibliography}
\end{document}